\documentclass[pre, twocolumn, showpacs, superscriptaddress]{revtex4-1}
\usepackage{amsmath}
\usepackage{mathrsfs} 
\usepackage{amsfonts}
\usepackage{graphicx}
\usepackage{subfigure}

\begin{document}
\title{Compressive Transition Path Sampling}
\author{Mehmet S\"uzen}
\email{mehmet.suzen@physics.org;suzen@fias.uni-frankfurt.de}
\date{\today}

\begin{abstract}
Algorithms for rare event complex systems simulations are proposed. Compressed Sensing (CS) 
has {\it revolutionized} our understanding of limits in signal recovery and has forced 
us to re-define Shannon-Nyquist sampling theorem for sparse recovery. A formalism to reconstruct 
trajectories and transition paths via CS is illustrated as proposed algorithms. The implication 
of under-sampling is quite important. This formalism could increase the tractable 
time-scales {\it immensely} for simulation of statistical mechanical systems and rare
 event simulations. While, long time-scales are known to be a major hurdle and a challenge 
for realistic complex simulations for rare events.  The outline of how to implement, 
test and possible challenges on the proposed approach are discussed in detail.
\end{abstract}
%
%
\maketitle

Simulation methods are now appearing as a standard tool to investigate structure and dynamics 
of complex systems. These methods rely on solving equations of motion, for deterministic or 
stochastic dynamics, needs to sample trajectories or set of moves over 
time \cite{allen1989computer,frenkel2002understanding,rapaport2004art}. Using these methods 
in rare events is shown to be a challenging task due to the presence of energy-barriers 
and meta-stable states, so special techniques should be used instead 
\cite{doltsinis3free,landau2000guide,newman1999monte,vanden2009some}, for example 
in studying transition states.

Analog signals can be sampled in a digital manner and the Shannon-Nyquist theorem 
\cite{shannon49, nyquist28} restricts how this can be achived in perfect manner. 
However, it is now known that under certain assumptions reconstructions can be  
achived with much less sampling, via \textit{compressed sensing} (CS) framework 
\cite{candes06a,donoho06a, eldar2012a}. 

Transition State Theory (TST) provides a theoretical framework to study barrier-crossing 
problems and rare events in complex systems 
\cite{eyring1935activated,wigner1938transition,horiuti1938statistical,haenggi1990reaction,
truhlar1996current,schenter2003generalized,doll05tst}. And TST is still an active area of 
research in chemical physics \cite{chong2017a,jung2017a}
A major concept introduced by TST is that a configuration of a complex system moves from a 
reactant state to a product state by navigating over saddle point of the potential energy 
surface i.e. a dividing surface, for example applied to isomerization dynamics 
\cite{chandler1978statistical}. Computing reaction rates over this saddle surface appears 
as a great challenge and attracts interest more then fifty 
years \cite{haenggi1990reaction,miller1998direct}. An important quantity in TST appears 
as reaction coordinate, an observable depending upon trajectory, which in most cases 
determined with intuition. This can be misleading in situation where slow varying variables 
are noting to do with reaction. Additionally, in many complex systems the very notion of 
transition state is obscured in higher dimensional space \cite{weinan2010transition}. To 
overcome these serious set back in TST, a set of novel approaches has pioneered by 
Pratt \cite{pratt1986statistical}, Transition Path Sampling (TSP) 
algorithms \cite{dellago1998transition,bolhuis1998sampling,bolhuis2002transition,
bolhuis2003transition,van2003novel,moroni2005efficient,chopra2008improved,
dellago2009transition,weinan2010transition} or Transition Path Theory (TPT) 
\cite{vanden2006transition,metzner7transition,weinan2006towards}, for example the 
string method \cite{ren2002string,weinan2005finite,weinan2010transition}. 
Instead of tracking transitions from a saddle point, TSP algorithms focus on transition 
paths i.e. pieces of trajectories which rare events occur. 
This approach developed much further to solve a realistic problem \cite{bolhuis2003transition},
for mathematically sound generalized framework \cite{bolhuis2002transition} and for 
meta-stable states \cite{rogal10a}.

Development of method(s) to study, transition pathways for rare events in complex systems
by using compressive sampling framework is shown in this article. This could be realized by devising
algorithm for sparse reconstruction of randomly under-sampled trajectories and phase-space regions.
Thereafter, implementation and testing of new algorithm(s) could be proceed on well studied physical
systems. If reconstructions of under-sampled trajectories and phase-space regions are realized, it
has quite significant effect on our ability to generate molecular motions by using 
much less {\it information}. This may allow us to simulate and investigate systems for much longer 
time-scales or transition problems having large reaction rates i.e. slow reactions. 

In Section \ref{str}, we have formalize how to reconstruct given tranjectory via undersampling,
in Section \ref{tpr} we outline a similar reconstruction scheme for transition paths are presented 
and in Section \ref{ci} challenges in implementation is discussed. An the last section, we summarize
an outlook. 

\section{Sparse Trajectory Reconstruction}
\label{str}
   Consider a trajectory sampled with equidistant time intervals which is represented 
   by a vector ${\bf x}(t) = ( {\bf p} (t),{\bf q}(t) )$ for $N$ component classical 
   system governed by Hamiltonian Dynamics. Sampled time points $t_{i}$ lies in the interval
   $[t_{1},t_{P}]$, where $i=1,...,P$, ${\bf p}_{j} (t_{i})$ and ${\bf q}_{j} (t_{i})$ are 
   written as momenta and coordinates at time $t_{i}=i \cdot \Delta t$ for particle $j=1,..,N$ 
   respectively. Hence a vector ${\bf x}(t) \in \mathbb{R}^{6N \cdot P}$ is defined as follows ;
      $${\bf x}(t) = ( {\bf p}_{1} (t_{1}), {\bf q}_{1} (t_{1}),...,$$ 
      $${\bf p}_{N} (t_{1}), {\bf q}_{N} (t_{1}),...,{\bf p}_{N} (t_{P}), {\bf q}_{N} (t_{P}),...{\bf p}_{N} (t_{P}), {\bf q}_{N} (t_{P}) ) ^{T} $$
   where position and momenta contain three components,
   ${\bf p}_{j} (t_{i})=(p_{j}^{x} (t_{i}),p_{j}^{y} (t_{i}),p_{j}^{z} (t_{i}))$ and 
   ${\bf q}_{j} (t_{i})=(q_{j}^{x} (t_{i}),q_{j}^{y} (t_{i}),q_{j}^{z} (t_{i}))$. At this point, 
   we asked the following question: Can we recover the same trajectory from a smaller 
   number of sampling points over time? The answer might be yes if we can follow up 
   CS framework presented in the previous section for a sparse signal recovery;

\begin{figure}
  \centering
  \includegraphics[angle=0,width=0.45\textwidth]{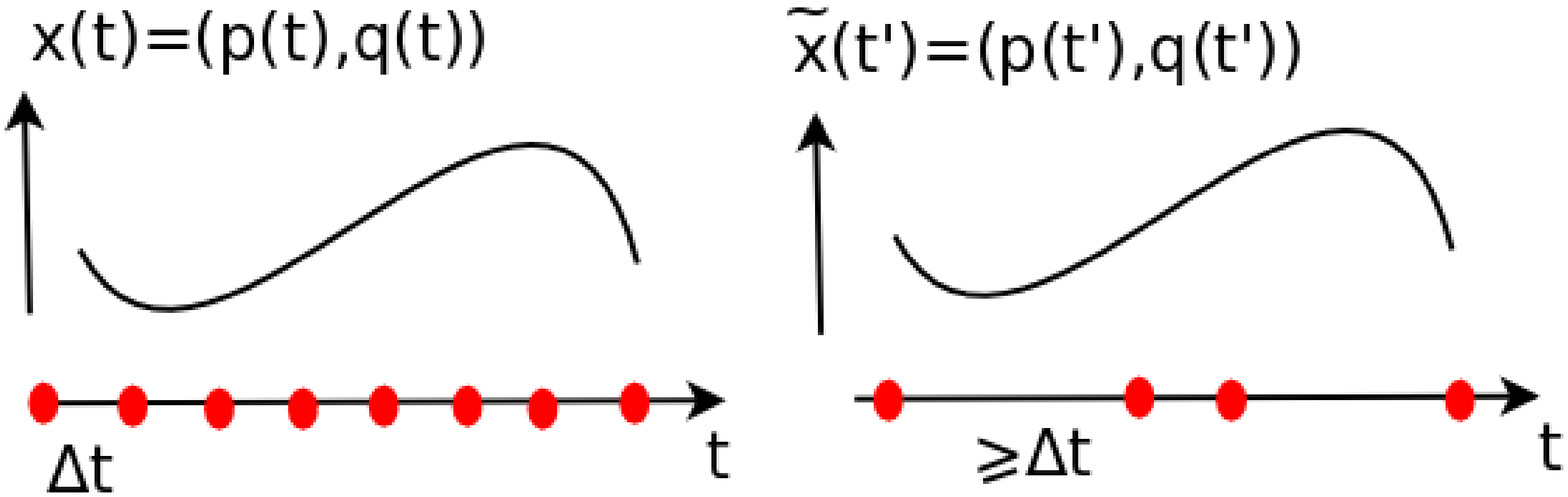}
  \caption{Conceptual sketch of a comparison between randomly sampled measurement 
           vector ${\bf \tilde{x} }$ (trajectory) and the {\it standard way} of 
           producing trajectory via regular samples in ${\bf x}$.}
   \label{xtilde}
\end{figure}

\begin{enumerate}
   \item The sparse representation of  ${\bf x} (t)$ via an orthogonal transformation $\mathscr{T}$ for example Discrete Fourier Transform or a wavelet bases can be written as 
     \begin{equation}
        {\bf x} (t)=\mathscr{T}  {\bf s } (t),
     \end{equation}
    ${\bf s } (t)$  being the sparse representation of the trajectory.
   \item A CS matrix ${\bf \Phi}$ is formed with an introduction of a { \it Gaussian random measurement 
         matrix} $\mathscr{G}$, ${\bf \Phi } = \mathscr{G} \mathscr{T}$, while it is known
         that random matrices are maximally incoherent to any bases. 
   \item  To be able to recover unknown trajectory randomly sampled measurements ${\bf \tilde{x} }(t')$ must be selected. Sampling realized with random time intervals which is represented by a vector ${\bf \tilde{x}}(t') = ( {\bf p} (t'),{\bf q}(t') )$ for $N$ component classical system governed by Hamiltonian Dynamics. Sampled points $t'_{l}$ lies in the interval $[t'_{1},t'_{Q}]$, where $l=1,...,Q$, ${\bf p}_{m} (t'_{l})$ and ${\bf q}_{m} (t'_{l})$ are written as momenta and coordinates at time $t'_{l}= t'_{l-1} + n \cdot \Delta t$ for particle $m=1,..,N$ respectively, and $n$ is a random number that generates next time-step randomly.  Hence a vector ${\bf \tilde{x} }(t') \in \mathbb{R}^{6N \cdot Q}$ is defined as follows;
    $$  {\bf \tilde{x} }(t') = ( {\bf p}_{1} (t'_{1}), {\bf q}_{1} (t'_{1}),...,$$ 
    $${\bf p}_{N} (t'_{1}), {\bf q}_{N} (t'_{1}),...,{\bf p}_{N} (t'_{Q}), {\bf q}_{N} (t'_{Q}),...{\bf p}_{N} (t'_{Q}), {\bf q}_{N} (t'_{Q}) ) ^{T}  $$
   where position and momenta contain three components ${\bf p}_{j} (t'_{l})=(p_{m}^{x} (t'_{l}),p_{m}^{y} (t'_{l}),p_{m}^{z} (t'_{l}))$ and ${\bf q}_{m} (t'_{l})=(q_{m}^{x} (t'_{l}),q_{m}^{y} (t'_{l}),q_{m}^{z} (t'_{l}))$.
     Comparison of two different sampling scheme is shown in Figure \ref{xtilde}.
   \item An optimization problem formulated as follows; 
        \begin{equation}
         min \; || \mathscr{T} {\bf s} ||_{1} \;\;\; s.t. \;\;\; {\bf \tilde{x} }={\bf \Phi} {\bf s},
        \end{equation}
         where unknown trajectory ${\bf x}$ will be recovered from this procedure.
\end{enumerate}

     There are some challenges in this procedure both from implementation and from physics 
     point of view which will be discussed later.

\begin{figure}
  \centering
  \includegraphics[angle=0,width=0.45\textwidth]{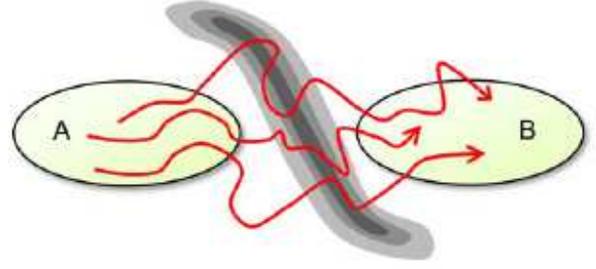}
  \caption{Reactive and product regions of phase-space A and B respectively. System
           remains in these regions much longer compare to rest of the 
           phase-space, taken from \cite{dellago2009transition}. }
   \label{tpsAB}
\end{figure}

\section{Transition Path Reconstructions}
\label{tpr}
   The basic machinery of Transition Path Ensemble is formulated by Dellago et. al. \cite{dellago1999calculation},
   the formulations can be varied in the literature \cite{weinan2010transition} but the basic idea is similar.

   The basic notion in describing transition paths is shown in Figure \ref{tpsAB} where a complex
   system undergoes a transition from region A to region B in the phase-space.
   These regions are stable in a sense that system stays considerably long.

   Consider a state at time $t$ which is explained with an instantenous trajectory of $N$ 
   particle system
   $$ {\bf x}_{i} (t) = ( {\bf p}_{1} (t_{i}), {\bf q}_{1} (t_{i}),...,{\bf p}_{N} (t_{i}), {\bf q}_{N} (t_{i}) ) .$$
   Introducing an {\it order parameter } $\lambda (x) $ may help us to identify region B, where product
   states are located,
   $$  {\bf x}_{i} (t)  \in B \; \; \; if \; \;  \lambda_{min}  \le \lambda (x) \le \lambda_{max} $$
   The distribution $P(\lambda,t)$ at time $t$ for trajectories starting in the region A at time $t=0$ 
   \begin{eqnarray}
      P(\lambda,t) & = & \int d x_{0} \rho(x_{0}) h_{A}(x_{0}) \delta \nonumber \\
                   & & \left[ \lambda - \lambda(x_{t})\right] \left( \int d x_{0} \rho(x_{0}) h_{A}(x_{0}) \right)^{-1}
     \label{ppp}
   \end{eqnarray}
    where $\rho(x_{0}$ is the equilibrium phase-space distribution,  $h_{A}$ and $h_{B}$ are characteristic
    functions which are either $1$ or $0$ depending upon if trajectory is inside the
    region A or B or not inside respectively and $delta$ is the usual Dirac delta-function. 
    The time correlation function  $C(t)$ then defined as follows
    \begin{equation}
      C(t) =  \int_{\lambda_{min}}^{\lambda_{max}} d \lambda \; P(\lambda,t) 
    \end{equation} 
      
     In order to compute $P(\lambda,t)$ we define $N-1$ overlapping regions $B[i]$ over the
     order-parameter space, where $B[0]=B$, such that 
     $$ {\bf x}_{t} \in B[i]  \Leftrightarrow  \lambda_{min}[i]  \le \lambda (x) \le \lambda_{max}[i] $$
     where index $i$ ranges $0 < i < N$ and region $B[i]$ must have an overlapping 
     windows with $B[i-1]$ and $B[i+1]$, so the probability of reactive trajectories for each region
     can be written 
     \begin{eqnarray}
       P(\lambda,t;i) & = & \int dx_{0} f_{AB[i]} (x_{0},t) \delta \nonumber \\ 
                      && \left[ \lambda - \lambda(x_{t})\right] \left( \int d x_{0} \rho(x_{0}) h_{A}(x_{0}) \right)^{-1},
     \end{eqnarray}
       where this equation is directly proportional to Eq. \ref{ppp}, $f_{AB[i]}$ is called {\it transition path ensemble}
       that describes all initial condition $x_{0}$ in region A leading to trajectories ending in $B[i]$ at time 
       $t$ ;
       $$ f_{AB[i]} = \rho(x_{0}) h_{A}(x_{0}) h_{B[i]}(x_{t}).$$
       One can compute time correlation function $C(t)$  (implies ability to compute reaction rates) 
       by matching histograms of $P(\lambda,t;i)$ in the overlapping regions to obtain $P(\lambda,t)$.
       Sampling this path ensemble was an intense research over the last decade. 

\subsection{Sparse Transition Path Ensemble}

     Recall the construction of transition path ensemble which has explained shortly. 
     Now, CS framework will be introduced in construction of transition path ensemble.
     The main idea is to construct $P(\lambda,t;i)$ histograms via CS framework. Consider
     the histograms as a vector ${\bf P}$ with regular samples $\Delta \lambda$ of $n$-bins.
      \begin{enumerate}
        \item The sparse representation of  ${\bf P} $ via an orthogonal transformation
              $\mathscr{T}$, for example Discrete Fourier Transform or a wavelet bases, can be written as
              \begin{equation}
                {\bf P} (\lambda,t)=\mathscr{T}  {\bf P_{s} } ,
              \end{equation}
              ${\bf P_{s} }$  being the sparse representation of the probability $P(\lambda,t;i)$.
        \item A CS matrix ${\bf \Phi}$ is formed with an introduction of a { \it Gaussian random measurement
         matrix} $\mathscr{G}$, ${\bf \Phi } = \mathscr{G} \mathscr{T}$, while it is known
         that random matrices are maximally incoherent to any bases.
        \item We define a measurement which is randomly under-sampled histogram of 
              probabilities ${\bf \tilde{P} } (\lambda,t;i)$, having randomly placed (random widths)
              $m$-bins, $m \ll n$ .
        \item An optimization problem formulated as follows;
        \begin{equation}
         min \; || \mathscr{T} {\bf P_{s}} ||_{1} \;\;\; s.t. \;\;\; {\bf \tilde{P} }={\bf \Phi} {\bf s},
        \end{equation}
         where unknown Probability ${\bf P}$ will be recovered from this procedure as well as correlation
         function and reaction rates as a consequence. The above procedure is called 
         {\it compressive transition path sampling }.
      \end{enumerate}
      The proposed method can be used with any of the the path sampling algorithms while the 
      {\it compression} is taken place in construction of probability histograms.
       
\section{Challenges in Implementation}
\label{ci}

Some challenges on implementing proposed formalism are discussed.
       \begin{enumerate}
         \item {\bf $\ell_{1}$ minimizer}: There are available minimizers written for 
          general purpose packages like matlab \cite{spgl07,CandRomb07}. For test purposes, mensioned
          packages would be sufficient, however for larger scale molecular systems distributed implementation
          is needed.
         \item {\bf Test systems for sparse trajectory construction} 
               One of the simplest system, Lennard-Jones liquid can be used \cite{smit1992phase} to 
               demonstrate sparse trajectory construction. For an initial test, only a randomly generated
               sub-set of obtained trajectory can be used i.e. retaining the physics of the trajectory 
               by using equily-spaced time-step. This means an {\it offline} analysis of the trajectory using
               random parts of it to reconstruct the original data. If this test is successful, the data (trajectory)
               obtained by randomly spaced time-steps can be tested, see challenges section.
         \item {\bf Test systems for sparse TPS} For inital test purposes, a model system studied previously 
               in the context of transition path sample \cite{dellago1999calculation}, which is 
               called Straub-Borkovec-Berne \cite{straub1988molecular}, can be utilized. Trajectories can be 
               generated via a standard coarse-grained codes such as LAMMPS \cite{plimpton2007lammps} or
               NAMD \cite{phillips2005scalable}. It is also an option to use smaller scale 
               snips of codes to make thinks much easier \cite{rapaport2004art} to have a compact tools.
               Further collection of test systems can also be used \cite{metzner2006illustration}.
         \item {\bf Realistic System} If initial test were successful enough a more realistic simulations
               can be tested, such as Protein folding \cite{bolhuis2003transition}.
         \item {\bf Physics of inverse problem } The equations presented for inverse reconstruction
          for sparse trajectory and sparse histograms in transition paths are based on generic 
          signal recovery. One may argue that the physics behind this approach is not {\it strong} enough. However
          the measurement vectors in both cased are indeed generated via physical process i.e. 
          molecular trajectories. For example in one pixel camera example of CS frame work 
          \cite{baraniuk07ln}, voltages generated by the lens is taken granted as measurements
          that are related to image, so there is no reason not to relate measured under-sampled 
          trajectories to sparse trajectories. But further justification of inverse problems proposed 
          in the previous sections must be developed in more rigorous mathematical terms, probably
          in the language of Hamiltonian Systems. Monte Carlo sampling was also proposed for
          inverse problems \cite{mosegaard1995monte}, this work maybe taken as a reference point.
         \item {\bf System size } In the proposed scheme whole trajectory evolution 
          of $N$-body system is dump into a single vector for sparse trajectory construction. This
          might be problematic for large systems with too many samples over time, for example 10000 
          particles with 3 ns simulations with 1 fs needs a storage of more then 
          3 million elements. However this problem 
          can be solved by introducing and iterative scheme for the minimizer that only needs to
          store adjacent sampling point i.e. time steps. 
         \item {\bf Using large time-steps} The real advantage of sparse recovery can be obtain when
          random large time-steps is used in producing molecular trajectories {\it on the fly}. However in that case, 
          the effect of large time-steps in the integrator and for the physics of the problem might 
          be in question. This problem studied in the literature extensively 
          \cite{winger2009using,fincham1986choice,macgowan1988large}.
       \end{enumerate}
  \section{Outlook}

We proposed a formalism to use CS in TPS, we can generate results from computer models of rare events 
{\it very fast}.  TPS is not only applicable to chemical reactions but on any complex system 
having a reaction mechanism, from one stable state to another, such as a power grid network into a 
failure state, a financial market from one state to an other, many more examples from complex networks 
can be given such as social networks. If an analogous concepts of trajectories and order parameters can be found 
for the mentioned complex systems.  Possible extensions of this formalism to stochastic noisy simulations is 
also possible where CS shown to work better in noisy environments.  A role of information content in 
transition path techniques can also be addressed. \\

\bibliographystyle{apsrev4-1}
\bibliography{msuzen}
\end{document}